\documentstyle[preprint,aps]{revtex}

\begin{document}

\draft

\title{Electron states in aone-dimensional random binary alloy}

\author{F.\ Dom\'{\i}nguez-Adame$^a$, I.\ G\'{o}mez$^a$, A.\ Avakyan$^b$, D.\
Sedrakyan$^c$, and A.\ Sedrakyan$^{b}$}
\address{$^a$GISC, Departamento de F\'{\i}sica de Materiales, Universidad\\
Complutense, E-28040 Madrid, Spain\\
$^b$Yerevan Physics Institute, Br.\ Alikhanian str.\ 2, Armenia\\
$^c$Yerevan State University, Armenia}
\date{\today}
\maketitle

\begin{abstract}
We present a model for alloys of compound semiconductors by introducing a
one-dimensional binary random system where impurities are placed in one
sublattice while host atoms lie on the other sublattice. The source of disorder
is the stochastic fluctuation of the impurity energy from site to site.
Although the system is one-dimensional and random, we demonstrate analytical
and numerically the existence of extended states in the neighborhood of a given
resonant energy, which match that of the host atoms.
\end{abstract}


\narrowtext

\section{Introduction}

Electron states in random systems have become an active research topic since
the generality of localization phenomena in one dimension (1D)~\cite{Mott61}. 
Although it is well established that almost any nonzero disorder causes
exponential localization of all eigenstates in 1D systems, regardless their
energy (see, e.~g., Ref.~\onlinecite{Ziman} and references therein), there
exist several exceptions. It is nowadays well known that extended states may
arise in random systems where disorder exhibits
short-range~\cite{Flores,Dunlap,Wu,Wu2,Wu3,Bovier,Wu4,Evan2,Datta0,Evan1,%
Flores2,PRBKP,JPA,Evan3,Datta,Diez} or long-range
correlations~\cite{Moura98,Izrailev99}. Spatial correlation means that random
variables are not independent within a given correlation length. Suppression of
localization by correlations was further put forward for the explanation of
high conductivity of doped polyaniline~\cite{Wu3} as well as transport
properties of random semiconductor superlattices~\cite{Bellani99}.

In this work, we report further progress along the lines in the preceding
paragraph. We turn ourselves to one of the pioneering works in the field,
namely the work of Wu and Phillips on polyaniline (see Ref.~\onlinecite{Wu3}
and references therein). These authors showed that polyaniline can be mapped
onto a random dimer model that has a set of extended states, originated by a
resonance of a single dimer defect (two neighbor sites with the same energy).
Electron states whose energy is close to this resonance turn out to be extended
(in the sense that their localization length is larger than the system size)
when dimers are placed at random in the 1D system. In this work we show that
{\em dimers\/} are not needed to observe extended states in 1D random systems
with short-range correlated disorder. To this end, we built up a simple model
of semiconductor binary alloy ---like ternary III-V compounds---. In these
alloys (say Al$_{x}$Ga$_{1-x}$As), the cation sublattice is occupied by the
same atoms (say As) while anions (say Al and Ga) are randomly distributed over
the other sublattice. We model these alloys by considering a 1D random binary
alloy with two species, referred to as A and B atoms hereafter. In order to
mimic the disorder present in the anion sublattice, we further assume that the
site energy of A atoms is randomly distributed from site to site while that of
B atoms is the same over the entire cation sublattice. As a major point, we
demonstrate the occurrence of extended states in the vicinity of the site
energy of B atoms  {\em in spite of the fact that the system is purely 1D and
random\/}. Thus, we conclude that dimers (or larger defects like trimers or
n-mers) are not required to observe extended states in 1D systems with
short-range correlated disorder.

\section{Model}

We consider a 1D binary system where A (B) atoms are placed at odd (even)
positions of the otherwise regular lattice, whose corresponding site
energies are $\epsilon_{2n-1}$ ($\epsilon_{2n}$) with $n=1,2,\ldots N$, $N$
being the number of unit cells of the alloy. The Schr\"{o}dinger equation
for stationary eigenstates $\psi_n(E)$ is 
\begin{equation}
(E-\epsilon_n)\psi_n +\psi_{n+1}+\psi_{n-1}=0, \quad n=1,2,\ldots,{\cal N},
\label{H}
\end{equation}
where the $E$ is the corresponding eigenenergy, $\epsilon_n$ is the site energy
and ${\cal N}\equiv 2N$ is the number of atoms in the system. According to our
model, site energy at even positions is the same and we can set
$\epsilon_{2n}=0$ without loss of generality. The source of disorder in this
model arise from the stochastic fluctuations of site energy at even positions.
We assume that $\{\epsilon_{2n-1}\}_{n=1}^{N}$ is a set of  {\em
uncorrelated\/} random Gaussian variables with mean value $v$ and  variance
$\sigma^2$. Hereafter $\sigma$ will be referred to as {\em degree of
disorder\/}. The joint distribution function of a realization of disorder is
represented by the direct product of single Gaussians. Thus 
\begin{equation}
\langle \epsilon_{2n} \rangle = v, \quad \quad \langle \epsilon_{2n}^2
\rangle = v^2 + \sigma^2.  \label{means}
\end{equation}

The Schr\"{o}dinger equation~(\ref{H}) can be written via the $2\times 2$
promotion-matrix $P_{n}$ as follows 
\begin{equation}
\left(
\begin{array}{l}
\psi_{n} \\ 
\psi_{n+1}
\end{array}
\right) = \left( 
\begin{array}{cc}
0 & 1 \\ 
-1 & -E+\epsilon_n
\end{array}
\right) \left(
\begin{array}{l}
\psi_{n-1} \\ 
\psi_{n}
\end{array}
\right) \equiv P_{n} \left(
\begin{array}{l}
\psi_{n-1} \\ 
\psi_{n}
\end{array}
\right).  \label{promotion}
\end{equation}
By iterating this equation we can relate $(\psi_{n},\psi_{n+1})$ and $%
(\psi_{0},\psi_{1})$ with $\psi_{0}\equiv 0$: 
\begin{equation}
\left(
\begin{array}{l}
\psi_{n} \\ 
\psi_{n+1}
\end{array}
\right) = \prod_{k=n}^{1} P_{k} \left(
\begin{array}{l}
\psi_{0} \\ 
\psi_{1}
\end{array}
\right) \equiv M_{n} \left(
\begin{array}{l}
\psi_{0} \\ 
\psi_{1}
\end{array}
\right),  \label{transfer}
\end{equation}
where $M_{n}$ is referred to as the transfer-matrix. We find most convenient to
deal with the promotion-matrix of the (diatomic) unit cell instead of that
corresponding to a single atom~(\ref{promotion}), namely $T_n \equiv
P_{2n}P_{2n-1}$. For real $E$ and $\epsilon_n$, the promotion-matrix $T_n$ can
be regarded as an element of the $SO(1,2)$ group, isomorphic to $SL(2,R)$. It
can be cast in the following form via the Pauli matrices $\sigma_\mu$ 
\begin{equation}
T_n =\Big[{E\over 2}\,(E -\epsilon_{2n-1})-1\Big]\,{\cal I}_{2} -
{E\over 2}\,(E -\epsilon_{2n-1})\sigma_3 +{\frac{\epsilon_{2n-1}}{2}}\sigma_1 
+ i\left(E-{\frac{\epsilon_{2n-1} }{2}}\right) \sigma_2,  
\label{T2}
\end{equation}
where ${\cal I}_{m}$ denotes the $m \times m$ unit matrix. It is easy to
demonstrate the following useful property $T_n^{-1}= \sigma_2
T_n^{\dag}\sigma_2$. The transfer-matrix of the entire system ($N$ unit cells)
is obtained as $M_{N}=\prod_{n=N}^{1}\,T_n$. Oseledec's theorem~\cite{O}
states that the following limiting matrix $\Gamma$ exists 
\begin{equation}
\Gamma= \lim_{{N} \rightarrow \infty} \left(M_{N}^{\dag}M_{N}^{}
\right)^{1/{2N}},  
\label{G}
\end{equation}
with eigenvalues $e^{\gamma}$. The Lyapunov exponent $\gamma$ is nothing but
the inverse of the localization length $\lambda^{-1}$, where $\lambda$ is given
in units of the length of the unit cell.

\section{Existence of extended states}

In order to find the localization length one should calculate the matrix 
$M_{N}^{\dag}M_{N}^{}$ for large $N$. We will perform this task following the
technique developed in Ref.~\cite{SS}. By using the formula for the
decomposition of the product of two spin-1/2 states into the direct sum of
scalar and spin-1 states, we have 
\begin{equation}
(T_j)^{\alpha}_{\alpha^{\prime}}(T_{j}^{-1})^{\beta^{\prime}}_{\beta}= {%
\frac{1 }{2}}(\delta)^{\alpha}_{\beta}
(\delta)^{\beta^{\prime}}_{\alpha^{\prime}}+ {\frac{1 }{2}}%
(\sigma^{\mu})^{\beta^{\prime}}_{\alpha^{\prime}}
\Lambda_{j}^{\mu\nu}(\sigma^{\nu})^{\alpha}_{\beta},  \label{TT1}
\end{equation}
where 
\begin{equation}
\Lambda_{j}^{\mu\nu}={\frac{1}{2}}\,{\rm Tr}\left(T_j\sigma^{\mu}T_{j}^{-1}
\sigma^{\nu}\right)  \label{L}
\end{equation}
is the spin-1 part. Multiplying the expression~(\ref{TT1}) by the left and
right by $\sigma_2$ we have 
\begin{equation}  \label{TT2}
(T_j)^{\alpha}_{\alpha^{\prime}}(T_{j}^{+})^{\beta^{\prime}}_{\beta}= 
{\frac{1}{2}}(\sigma_2)^{\alpha}_{\beta}
(\sigma_2)^{\beta^{\prime}}_{\alpha^{\prime}}+ {\frac{1 }{2}}(\sigma^{\mu}
\sigma_2)^{\beta^{\prime}}_{\alpha^{\prime}}
\Lambda_{j}^{\mu\nu}(\sigma^{\nu} \sigma_2)^{\alpha}_{\beta}.
\end{equation}

Now we should take into account the disorder and calculate the average of 
$\Gamma$ by random distribution of $\epsilon_{2n-1}$ at odd sites 
\begin{equation}
\langle\Gamma \rangle ={\frac{1 }{2}} \sigma_2 \otimes \sigma_2 + 
{\frac{1}{2}} (\sigma^{\mu} \sigma_2)\otimes (\sigma^{\nu} \sigma_2)
\left(\prod_{j=1}^{N} \langle\Lambda_{j}\rangle\right)^{\mu \nu},
\label{Gamma}
\end{equation}
where $\Lambda_j$ is defined by~(\ref{L}). According to Oseledec's 
theorem~\cite{O}, the Lyapunov exponent and, correspondingly, the localization
length will be given by 
\begin{equation}
\lambda^{-1}= \log[\xi(E)],  \label{CL}
\end{equation}
where $\xi(E)$ is the closest to unity eigenvalue of $\langle
\Lambda_j\rangle$, whose elements are 
\begin{eqnarray}
\langle\Lambda^{11}_{j}\rangle &=&{\frac{1 }{2}}\,
\left(2+E^4+4Ev-2E^3v-\sigma^2-v^2+(\sigma^2+v^2-4)E^2\right)  \nonumber \\
\langle\Lambda^{12}_{j}\rangle &=&-{\frac{i }{2}}\,\left(E^4-2E^3v
+\sigma^2+v^2+(\sigma^2+v^2-2)E^2\right)  \nonumber \\
\langle\Lambda^{13}_{j}\rangle &=&-E^3-v+2E^2v-(\sigma^2+v^2-2)E  \nonumber
\\
\langle\Lambda^{21}_{j}\rangle &=&{\frac{i }{2}}\,\left(E^4+4Ev-2E^3v
-\sigma^2-v^2+(\sigma^2+v^2-2)E^2\right)  \nonumber \\
\langle\Lambda^{22}_{j}\rangle &=&{\frac{1 }{2}}\,\left(2 + E^4 -2 E^3 v
+\sigma^2+v^2+ (\sigma^2+v^2)E^2\right)  \nonumber \\
\langle\Lambda^{23}_{j}\rangle
&=&-i\,\left(E^3+v-2E^2v+(\sigma^2+v^2)E\right)  \nonumber \\
\langle\Lambda^{31}_{j}\rangle &=&-2E+E^3+v-E^2v  \nonumber \\
\langle\Lambda^{32}_{j}\rangle &=&-i\,\left(E^3-v-E^2v\right)  \nonumber \\
\langle\Lambda^{33}_{j}\rangle &=& 1-2E^2+2Ev.  
\label{L3}
\end{eqnarray}

Delocalized states have an infinite localization length and, therefore, at some
particular energy $E$, the matrix $\langle\Lambda_{j}\rangle$ should have an
eigenvalue equal to one. Hence we obtain the following condition for obtaining
delocalized states 
\begin{equation}  \label{D}
\det\Bigg[{\cal I}_3-\langle\Lambda_{j}\rangle\Bigg]=-2\sigma^2E^2 = 0.
\end{equation}
As we see, there is a delocalized state at $E=0$. One can calculate the
localization length~(\ref{CL}) and expand it around $E=0$. For $v>0$ one
gets 
\begin{equation}
\lambda^{-1}= \left\{
\begin{array}{ll}
{\frac{\displaystyle \sigma^2 }{\displaystyle 4v}}\,E + {\cal O}(E^2), & E<0,\\ 
2v^{1/2}E^{1/2}+{\cal O}(E^{3/2}) , & E>0. 
\label{F}
\end{array}
\right.
\end{equation}
Notice that the localization length is asymmetric around the energy of the
extended state since it scales as $\sim E$ at the left and $\sim E^{1/2}$ at
the right. The situation is just the opposite for $v<0$. Remarkably, the
prefactor at the left depends on the degree of disorder of the alloy but
becomes independent of disorder at the right.

When the degree of disorder vanishes ($\sigma=0$), the alloy is simply a
diatomic periodic chain with site energies $0$ and $v$ in each unit cell and,
consequently, there are two allowed bands. The lower band ranges from 
$v/2-\left[(v/2)^2+4\right]^{1/2}$ up to $0$ while the upper band ranges from 
$v$ up to $v/2+\left[(v/2)^2+4\right]^{1/2}$ for $v>0$. Obviously all
eigenstates are Bloch functions and spread over the entire chain. Localization
occurs as soon as a small degree of disorder is introduced in the system. But,
according to our previous results, the eigenstate with $E=0$ remains extended.
This is clearly seen in Fig.~\ref{fig1}, where the inverse of the localization
length obtained from~(\ref{CL}) is plotted against $E$ for $v=1$ and different
degrees of disorder $\sigma$. In all cases the inverse of the localization
length is nonzero except at $E=0$, where  $\lambda^{-1}=0$. This suggests the
occurrence of a delocalized states at  $E=0$.

\section{Numerical results}

To confirm the above analytical results we have also numerically diagonalized
the Schr\"odinger equation~(\ref{H}). We will mainly focus our attention on the
normalized density of states $\rho (E)$ and on the degree of localization
(inverse participation ratio, IPR) for the states at energy  $E$. They are
defined respectively as follows~\cite{Fidder91} 
\begin{mathletters}
\begin{equation}
\rho (E) = {\frac{1}{{\cal N}}}\Biggl\langle\sum_k \left({\frac{1}{R}}
\right)\theta\Bigg[{\frac{R}{2}}-|E - E_k|\Bigg]\Biggr\rangle,  \label{rho}
\end{equation}
\begin{equation}
L(E) = {\frac{1}{{\cal N}\rho(E)}}\Biggl\langle\sum_k \left({\frac{1}{R}}
\right)\theta\Bigg[{\frac{R}{2}}-|E - E_k|\Bigg] \Biggl(\sum_{n=1}^{{\cal N}} 
a_{kn}^4\Biggr)\Biggr\rangle,  \label{IPR}
\end{equation}
\end{mathletters}
where the angular brackets indicate an average over an ensemble of disordered
linear chains and the $a_{kn}$ is the eigenvector of~(\ref{H}) corresponding to
the eigenvalue $E_k$ with $k=1,2,\ldots,{\cal N}$. Here $R$ is the spectral
resolution and $\theta$ is the Heaviside step function. The IPR behaves like
$1/{\cal N}$ for delocalized states spreading uniformly over the entire system
on increasing ${\cal N}$. In particular, the IPR can be exactly computed for
the eigenstates of the periodic lattices. In doing so we obtain the expected
behavior for ${\cal N}\to \infty$. On the contrary, localized states exhibit
much higher values. In the extreme case, when the eigestate is localized at a
single site, the IPR becomes unity.

We have fixed $v=1$ and studied several values of the degree of disorder 
$\sigma$, ranging from $0.1$ up to $1.0$. The highest value of the degree of
disorder considered in the present work means that the typical fluctuations of
the site energy is of the order of the nearest neighbor coupling. The maximum
number of atoms in the chain was ${\cal N}=1000$ ($500$ unit cells) although
larger systems were also studied to check that our main results are independent
of the size. The results comprise the average over $100$ realizations of the
disorder for each given pair of parameters $v$ and  $\sigma$. The spectral
resolution was $R=4\times 10^{-3}$.

Let us comment the results we have obtained numerically. Figure~\ref{fig2}
shows the DOS for two different values of the degree of disorder ($\sigma=0.1 $
and $\sigma=1.0$) when the system size is ${\cal N}=1000$ and $v=1$. The DOS
presents the usual U-shape within the bands when the degree of disorder is
small. The singularities at the edge of the allowed bands are smeared out on
increasing the degree of disorder except at $E=0$, where the divergence remains
even for the largest degree of disorder ($\sigma=1.0$). This result provides
further evidence that the states at $E=0$ are delocalized.

The degree of localization (IPR) presents an overall increase when the degree
of disorder increases, meaning that the larger the degree of disorder, the
smaller the localization length. This is clearly observed in Fig.~\ref{fig3},
where we show the IPR as a function of the energy for the same parameters of
Fig.~\ref{fig1}. However, the increase of the IPR strongly depends on the
energy, being more pronounced close to the center of both allowed bands.
Interestingly, the IPR at $E=0$ becomes independent on the degree of disorder
although depends on the system size, as expected. This peculiarity manifest the
delocalized character of the states at this spectral region. Finally, notice
the good correspondence between Fig.~\ref{fig1} (analytical results) and
Fig.~\ref{fig3} (numerical results).

\section{Conclusions}

In this paper we have considered electron dynamics in a one-dimensional model
of binary alloy where disorder lies in one of the two sublattices. Although the
system is purely one-dimensional and random, we have demonstrated analytically
the existence of delocalized states close to a resonant energy, which match
that of the atoms of the other sublattice. Numerical results from the
evaluation of the DOS and IPR (degree of localization) strongly suggest that
there exist many states close to the resonant energy that remain unscattered in
finite systems.

\acknowledgments

Work at Madrid is supported by CAM under Project~07N/0034/98. Work at Yerevan
is supported by INTAS grant.

\begin{figure}[tbp]
\caption{Inverse of the localization length as a function of energy when $v=1
$ and $\sigma=0.1$ (dashed line), $\sigma=0.5$ (solid line) and $\sigma=1.0$
(dotted line). Notice that $\lambda \to \infty$ at $E=0$.}
\label{fig1}
\end{figure}

\begin{figure}[tbp]
\caption{Density of states as a function of energy when the lattice size is $%
{\cal N}=1000$, $v=1$ and the degree of disorder is $\sigma=0.1$ (dashed
line) and $\sigma=1.0$ (dotted line).}
\label{fig2}
\end{figure}

\begin{figure}[tbp]
\caption{Inverse participation ratio for the same cases shown in Fig.~{\ref
{fig1}}. Notice the overall increase on increasing the degree of disorder
except at $E=0$.}
\label{fig3}
\end{figure}

\end{document}